# Phonons in Twisted Bilayer Graphene

Alexandr I. Cocemasov[1], Denis L. Nika[1,2,*] and Alexander A. Balandin[2,3,*]

[1]E. Pokatilov Laboratory of Physics and Engineering of Nanomaterials, Department of Theoretical Physics, Moldova State University, Chisinau, MD-2009, Republic of Moldova

[2]Nano-Device Laboratory, Department of Electrical Engineering, Bourns College of Engineering, University of California – Riverside, Riverside, California 92521 U.S.A.

[3]Materials Science and Engineering Program, Bourns College of Engineering, University of California – Riverside, Riverside, California 92521 U.S.A.

**Abstract**

We theoretically investigated phonon dispersion in AA-stacked, AB-stacked and twisted bilayer graphene with various rotation angles. The calculations were performed using the Born-von-Karman model for the intra-layer atomic interactions and the Lennard-Jones potential for the inter-layer interactions. It was found that the stacking order affects the out-of-plane acoustic phonon modes the most. The difference in the phonon densities of states in the twisted bilayer graphene and in AA- or AB-stacked bilayer graphene appears in the phonon frequencies range 90 – 110 cm$^{-1}$. Twisting bilayer graphene leads to emergence of new phonon branches – termed *entangled phonons* – which originate from mixing of phonon modes from different high-symmetry directions in the Brillouin zone. The frequencies of the entangled phonon depend strongly on the rotation angle and can be used for non-contact identification of the twist angles in graphene samples. The obtained results and tabulated frequencies of phonons in twisted bilayer graphene are important for interpretation of experimental Raman data and determining thermal conductivity of these materials systems.

* Corresponding authors: (DLN) dnica@ee.ucr.edu and (AAB) balandin@ee.ucr.edu

## I. Introduction

Two-dimensional sheet of $sp^2$ carbon atoms – *graphene* – is a promising material for future electronics because of its unique electrical [1-2], thermal [3-4] and optical [5-6] properties. The high electrical [1-2] and thermal [3-4] conductivities of graphene are crucial for its proposed applications in field-effect transistors [7], sensors [8], solar cells [9], resonators [10] and thermal management of ultra-large scale integrated circuits and high-power-density devices [4, 11-13]. In recent years the interest of the physics community has been shifting toward investigation of the twisted few-layer graphene (T-FLG) systems. When two graphene layers are placed on top of each other they can form a Moire pattern [14-16]. In this case, one layer is rotated relative to another layer by an arbitrary angle. The synthesis of T-FLG was experimentally demonstrated using the chemical vapor deposition (CVD), mechanical exfoliation or growth on the carbon terminated SiC surface [15-19]. Although twisting only weakly affects the interlayer interaction, it breaks symmetry of the Bernal-stacking resulting in an intriguing dependence of the electronic and phonon properties on the rotation angle (RA).

The electronic structure of the twisted bilayer graphene (T-BLG) with relatively small RAs was theoretically studied using both the continuum approach [14] and the density functional theory (DFT) [20-21]. It was demonstrated that the low-energy electron dispersion in T-BLG is linear as in a single layer graphene (SLG) but with the reduced Fermi velocity, especially for small RAs [14]. Two independent DFT studies [20-21] have shown that T-FLG systems possess the massless Fermion carrier property with the same Fermi velocity as in SLG both for incommensurate [20] and commensurate [21] graphene layers. Experimentally, the specifics of the electronic transport in T-FLG were investigated using the surface X-ray diffraction [17], scanning tunneling microscopy (STM) [17-18] and Raman spectroscopy [15]. It was observed that twisted multilayer graphene grown on the carbon terminated face of 4H-SiC reveals SLG electronic properties [17]. More recently, this observation was confirmed by an independent Raman spectroscopy study [15]. On the other hand, it was concluded on the basis of STM investigation [18] that the electronic properties in T-FLG are indistinguishable from those in SLG only for the large rotational angles > $20^0$ while for the small RAs the Dirac cone approximation breaks done even in the vicinity of *K*-point of Brillouin zone (BZ). While the electronic properties of T-FLG have been intensively investigated both theoretically and experimentally [14-23] the phonon properties of T-FLG remain largely unexplored.

The experimental data for phonons in T-FLG is expected to come mostly from the Raman spectroscopy, which proved to be a powerful tool for understanding phonons in SLG and FLG [24-34]. The earlier Raman studies of graphene [24-28] were focused on analysis of *G*-, *D*- and *2D*-band in graphene: peaks spectral position, shape and their dependence on the number of atomic planes in the samples. The *G*-band near 1485 $cm^{-1}$ is the first-order Raman peak in graphene related to scattering of in-plane transverse (TO) or longitudinal (LO) phonon of BZ $\Gamma$-point [24-27]. The pronounced *2D*-band in the range of 2500 – 2800 $cm^{-1}$ is the second-order Raman peak, associated with scattering of two TO phonons around *K*-point of BZ [24-27]. Ultraviolet Raman measurements provided additional data for determining the number of atomic layers in FLG due to different ratios of *G* and *2D* peaks [28]. The temperature dependence of Raman peaks was instrumental for the first measurement of the thermal conductivity of graphene [3-4, 11-13, 29]. More recent Raman studies revealed several new low-energy peaks in the spectra of FLG, which were attributed to $\Gamma$-point in-plane and out-of-plane acoustic phonon modes with non-zero energy [30-33]. Since these modes are sensitive both to the number of atomic layers and inter-layer coupling a possibility to use them for the non-contact characterization of the stacking order in FLG and T-FLG was proposed [32]. The new double-resonant peaks in the Raman spectra at $\omega$ ~ 1625 $cm^{-1}$, referred to as *R'* [16, 23], and $\omega$ ~ 1375 $cm^{-1}$, referred to as *R* [22, 23], were also observed in the twisted bilayer graphene. It was suggested that, depending on the rotational angle, the phonons with different wave vectors can participate in the processes leading to the appearance of these peaks [16, 22-23].

In the studies of the Raman processes in twisted graphene the phonon energies were treated as in non-twisted FLG [16, 23, 34]. However, the change in the stacking order and BZ size of T-FLG should affect their phonon properties. This assumption is confirmed by a recent theoretical study of the phonon modes in the nanometer-scale circular T-BLG sample of a finite radius in the range from 0.5 to 3 nm [35]. Using the Brenner potential for the intra-layer interactions and the Lennard-Jones potential for the inter-layer interaction, the authors found significant frequency shifts for ZO mode depending on the rotation angle [35]. This means that there is a strong need for an accurate theory and computation of the phonon energy, dispersion and density of states (DOS) in T-BLG for the purpose of interpretation of experimental data.

Here we present the first consistent theoretical study of the phonon modes in formally-infinite twisted bilayer graphene using the Born-von-Karman model of the lattice dynamics for in-plane atomic coupling and the Lennard-Jones potential for inter-layer atomic coupling. We do not limit the size of T-BLG and carefully take into account periodicity of the otationally-dependent unit cells. The rest of the paper is organized as follows. In Section II, we present our theoretical approach for calculation of the phonon modes in T-BLG. In Sections III and IV, we discuss our results, provide phonon frequency data for different rotation angles of T-BLG and give our conclusions.

## II. Theory of Phonons in Twisted Bilayer Graphene

We consider two parallel graphene sheets, referred to as the "bottom" and "top". We choose the initial configuration the same as for AA-stacked bilayer graphene (AA-BLG) and place the rotation center at the hexagon center. Then we rotate the "top" sheet relative to the "bottom" one in the graphene plane by an angle $\theta$ to obtain the twisted bilayer graphene. In a chosen rotation scheme, T-BLGs with the rotation angles $\theta = 0^0$ and $\theta = 60°$ correspond to the AA-stacked BLG, while T-BLGs with $\theta = 30° - \alpha$ and $\theta = 30° + \alpha$ are identical, where angle $\alpha \in [0, 30^0]$. It limits our consideration to twisting angles $\theta$ between 0 and 30 degrees.

Since the lattice dynamics approach is based on the periodicity of the crystal lattice, we, first, define a periodic (commensurate) atomic configuration of T-BLG. The angles of the commensurate rotations are given by the expression [14, 35-36]: $\cos\theta(m,n) = (3m^2 + 3mn + n^2/2)/(3m^2 + 3mn + n^2)$, where $m$ and $n$ are the positive integer numbers. If $n$ is not divisible by 3 then the basis vectors of Bravais lattice $\vec{t}_1$ and $\vec{t}_2$ for the commensurate T-BLG are given by the following expression:

$$\begin{pmatrix} \vec{t}_1 \\ \vec{t}_2 \end{pmatrix} = \begin{pmatrix} m & m+n \\ -(m+n) & 2m+n \end{pmatrix} \begin{pmatrix} \vec{a}_1 \\ \vec{a}_2 \end{pmatrix}, \quad (1)$$

where $\vec{a}_1 = (3a/2, -\sqrt{3}a/2)$ and $\vec{a}_2 = (3a/2, \sqrt{3}a/2)$ are the basis vectors of Bravais lattice for the single layer graphene. The smallest carbon-carbon distance is $a$=0.142 nm. The number of atoms in the commensurate cell is equal to the ratio between the unit cell volumes of the rotated and unrotated cells multiplied by a number of atoms in the unrotated cell:

$$N = 4\frac{\left|\left[\vec{t}_1 \times \vec{t}_2\right]\cdot \vec{z}\right|}{\left|\left[\vec{a}_1 \times \vec{a}_2\right]\cdot \vec{z}\right|}, \tag{2}$$

where $\vec{z}$ is the unitary vector normal to the graphene plane. Substituting Eqs. (1) into Eq. (2) one can obtain for $N$:

$$N = 4\left(\left(m+n\right)^2 + m\left(2m+n\right)\right). \tag{3}$$

Eq. (3) determines the number of atoms in the unit cell of T-BLG for a given pair of integers $(m,n)$. In Fig. 1 (a) we show a schematic view of the lattice structure of T-BLG with $\theta(1,1) = 21.8°$. In this case, the unit cell contains the smallest possible number of carbon atoms $N$ = 28. The T-BLG unit cells with other $\theta$ are larger. For instance, the unit cell with $\theta(2,1) = 13.7°$ contains 76 atoms.

We denote the basis vectors of the reciprocal lattice in BLG as $\vec{b}_1$ and $\vec{b}_2$, while the basis vectors of the reciprocal lattice in T-BLG as $\vec{g}_1$ and $\vec{g}_2$. The reciprocal vectors $\vec{g}_1$ and $\vec{g}_2$ are related to the real space commensurate vectors $\vec{t}_1$ and $\vec{t}_2$: $\vec{g}_1 = 2\pi\frac{\left[\vec{t}_2 \times \vec{z}\right]}{\left|\left[\vec{t}_1 \times \vec{t}_2\right]\cdot \vec{z}\right|}$; $\vec{g}_2 = 2\pi\frac{\left[\vec{z} \times \vec{t}_1\right]}{\left|\left[\vec{t}_1 \times \vec{t}_2\right]\cdot \vec{z}\right|}$. The same expressions can be written for the unrotated case. Using these relationships and Eqs. (1) one can express the vectors $\vec{g}_1$ and $\vec{g}_2$ through the vectors $\vec{b}_1$ and $\vec{b}_2$:

$$\begin{pmatrix} \vec{g}_1 \\ \vec{g}_2 \end{pmatrix} = \frac{1}{\left(m+n\right)^2 + m\left(2m+n\right)} \times \begin{pmatrix} 2m+n & m+n \\ -(m+n) & m \end{pmatrix} \begin{pmatrix} \vec{b}_1 \\ \vec{b}_2 \end{pmatrix} \tag{4}$$

Thus, using Eqs. (4) we are able to construct the Brillouin zone of the T-BLG with an arbitrary angle of rotation $\theta(m,n)$. The BZ of the T-BLG with $\theta(1,1) = 21.8°$ as well as of the rotated ("top") and unrotated ("bottom") single-layer graphene sheets which constitute the BLG is shown in Fig. 1 (b) by red, blue and black lines correspondingly.

**<Figure 1 (a, b)>**

The equations of motion for BLG and T-BLG atoms have the form:

$$\omega^2(\vec{q})U^i_\alpha(\vec{q}) = \sum_{\beta=1}^{3}\sum_{s=1}^{N_{sph}}\sum_{j=1}^{N_s} D^{ij}_{\alpha\beta}(s;\vec{q})U^j_\beta(\vec{q});\ \alpha = 1,2,3;\ i = 1,2,...,N. \quad (5)$$

Here $\omega$ is the phonon frequency, $\vec{q}$ is the two-dimensional phonon wave vector and $D^{ij}_{\alpha\beta}$ are the dynamic matrix coefficients:

$$D^{ij}_{\alpha\beta}(s;\vec{q}) = \frac{1}{m}\Phi^{ij}_{\alpha\beta}(s)\exp\left(\mathbf{i}\vec{h}^{ij}\vec{q}\right). \quad (6)$$

In Eqs. (5-6) $s$ denotes the nearest atomic spheres of the atom *i; j* denotes the atoms of the atomic sphere *s*; $\alpha,\beta$ designate the Cartesian coordinates components, $m$ is the mass of a carbon atom, $\Phi^{ij}_{\alpha\beta}(s)$ are the interatomic force constants describing the interaction between an atom $j$ and an atom *i*, located at the center of the atomic spheres, and $\vec{h}^{ij} = \vec{r}_j - \vec{r}_i$, where $\vec{r}_i$ [$\vec{r}_j$] is the radius vector of the atom *i*[*j*]. The components of the dynamic matrix are determined by the interatomic force constants. For this reason, a proper choice of the interatomic potential is crucial for obtaining the correct phonon energies. In this work, we apply the Born-von-Karman (BvK) model for description of the carbon-carbon intra-layer interaction. This model demonstrated a good agreement between theoretical and experimental phonon dispersions of bulk graphite [37-39].

In the case of the intra-layer coupling the hexagonal symmetry of the interatomic interaction is preserved for different $\theta$. Therefore, in Eq. (6) the force constant matrices and the coordination vectors for the "bottom" (unrotated) sheet will coincide with the single-layer graphene case. For the "top" (rotated) sheet these matrices and vectors can be found by applying a corresponding rotation operator. In our BvK model for the intra-layer interaction we take into account four nearest neighbor atoms $N_{sph}$ = 4. A schematic of the coordination spheres for the unrotated graphene sheet is presented in Fig. 2. A single-layer graphene has 2 atoms in the unit cell, which we denoted in Fig. 2 as "blue" and "red" circles. The vectors $\vec{h}^{ij}$ of atoms from the four atomic spheres of a "blue" atom are listed in Table I. Note that *z*th component of all atoms is equal to 0 and thus is omitted in the notations. The vectors $\vec{h}^{ij}$ of the atoms from the four atomic spheres of a “red” atom can be found by the rotation of the corresponding vector from Table I by the angle $\pi$ around the (0, 0)-point.

**<Table I>**

**<Figure 2>**

The force constant matrix describing the interaction of an atom with its $n^{\text{th}}$-nearest neighbor in graphene has the form: $\Phi^{(n)} = -\begin{pmatrix} \alpha^{(n)} & 0 & 0 \\ 0 & \beta^{(n)} & 0 \\ 0 & 0 & \gamma^{(n)} \end{pmatrix}$. The force constant matrices for all atoms from four neighbor atomic spheres can be found from $\Phi^{(n)}$ using the rotations around $Z$-axis by an angle $\varphi$ in a clock-wise direction $R_\varphi = \begin{pmatrix} \cos\varphi & \sin\varphi & 0 \\ -\sin\varphi & \cos\varphi & 0 \\ 0 & 0 & 1 \end{pmatrix}$ and the reflections in $XZ$-plane: $\sigma_y = \begin{pmatrix} 1 & 0 & 0 \\ 0 & -1 & 0 \\ 0 & 0 & 1 \end{pmatrix}$. In Table II we present a list of corresponding symmetry operations which map the matrix $\Phi^{(n)}$ into force constant matrices for all atoms from the first four atomic spheres.

**<Table II>**

The intra-layer interaction for each atomic sphere is described by three independent force constants: $\alpha_s$, $\beta_s$ and $\gamma_s$. In BvK model these constants have a clear physical meaning. A displacement of an atom induces a force towards its $n^{\text{th}}$ neighbor. The force constant $\alpha$ describes the longitudinal component of the force while constants $\beta$ and $\gamma$ describe the in-plane and out-of-plane transverse components, respectively. In our calculations, we used the force constants determined in Ref. [37] from a comparison between the theoretical and experimental phonon dispersion curves of bulk graphite. The numerical values of these constants are provided in Table III.

We describe the inter-layer interactions with the Lennard-Jones (L-J) potential $V(r) = 4\varepsilon\left((\sigma/r)^{12} - (\sigma/r)^{6}\right)$, with $\varepsilon$ = 4.6 meV, $\sigma$ = 0.3276 nm and 0.5 nm cutoff distance. The parameters $\varepsilon$ and $\sigma$ were taken from Ref. [40]. They reproduce the experimental values of the interlayer space and phonon dispersion along the $\Gamma - A$ direction of bulk graphite. In the case of the inter-layer coupling, the atomic configuration and force constant matrices are dependent on the rotation angle $\theta$:

$$\Phi_{\alpha\beta}^{ij}(n;\theta) = -\delta(r(\theta)) \times \frac{r_{\alpha}(n;\theta) r_{\beta}(n;\theta)}{\left|\vec{r}(n;\theta)\right|^{2}}, \qquad (6)$$

where $\delta(r(\theta)) = 4\varepsilon\left(\frac{156\sigma^{12}}{r^{14}(\theta)} - \frac{42\sigma^{6}}{r^{8}(\theta)}\right)$ is the force constant of the inter-layer coupling, $r(\theta)$ is the distance between the interacting atoms from a given atomic configuration corresponding to angle $\theta$.

**<Table III>**

## III. Phonon Dispersion in Twisted Bilayer Graphene

In the last few years significant efforts have been undertaken for investigation of the physical properties of AB-stacked FLG, which has higher stability in comparison with AA-stacked FLG [11-13, 27-31]. However, successful preparation of AA-stacked FLG has also been reported [41-42]. It was indicated in one report that synthesized bilayer graphene often exhibits AA-stacking, which makes difficult distinguishing it from SLG [42]. Theoretical investigation of electronic, magneto-optical, Raman and infrared properties in AA-stacked BLG revealed their strong difference from those of AB-stacked BLG (AB-BLG) [43-45]. Although the main goal of this work is to investigate the evolution of the phonon modes in T-BLG, we first compare the phonon dispersions in SLG, AA-BLG and AB-BLG. The latter is done in order to both validate our theoretical approach and to explore the difference in the phonon energy spectra between AA-BLG and AB-BLG.

The phonon dispersions in SLG, AA-BLG, AB-BLG and T-BLG with the rotation angles $\theta = 21.8°$ and $\theta = 13.7°$ are shown in Figs. 3 and 4. In Fig. 3 we plot the dispersions along

*Γ-K* direction in Brillouin zone of SLG, BLG or T-BLG, correspondingly. The phonon dispersion near *Γ*- and *K*-points is shown in Fig. 4. The phonon frequencies were calculated from Eqs. (5) for each phonon wave number $q$ from the interval 0 to $q_{\max}(\theta)$, where $q_{\max}(\theta) = 2q_{\max}(\theta = 0)\sin(\theta/2) = 8\pi\sin(\theta/2)/(3\sqrt{3}a)$. For SLG and BLG $q_{\max} = q_{\max}(\theta = 0) = 4\pi/(3\sqrt{3}a)$. The directions in BZ of T-BLG depend strongly on the rotational angle and do not coincide with the directions in BZ of SLG or BLG. As shown in Fig. 1 (b), the *Γ-K* direction in BZ of T-BLG is rotated relative to that in BZ of BLG. Therefore, the phonon curves in Fig. 3 (a-d) are shown for different directions in BZ of BLG. However, the *Γ*- and *K*-points in BZ of T-BLG correspond to that in BZ of BLG and the change of the phonon modes in these points is a direct effect of the twisting.

The unit cell of SLG consists of two atoms, therefore there are six phonon branches in SLG (see Fig. 3 (a)): the out-of-plane transverse acoustic (ZA), out-of-plane transverse optic (ZO), in-plane longitudinal acoustic (LA), in-plane transverse acoustic (TA), in-plane longitudinal optic (LO) and in-plane transverse optic (TO). In BLG, there are four atoms in the unit cell and the number of phonon branches is doubled (see Fig. 3 (b)). We denoted the polarizations of these pairs of branches as follows: $LA_1/LA_2$, $TA_1/TA_2$, $ZA_1/ZA_2$, $LO_1/LO_2$, $TO_1/TO_2$ and $ZO_1/ZO_2$. These modes, in general, can be understood as the "bilayer" analogs of LA, TA, ZA, LO, TO and ZO polarizations of SLG. The energy difference $\Delta$ between the phonon branches in the pairs is small due to the weak van der Waals coupling. It attains its maximum value $\Delta_{\max}$ at the BZ center: $\Delta_{\max}(\mathrm{LA}) = \Delta_{\max}(\mathrm{TA}) = 13.4\ \mathrm{cm}^{-1}$, $\Delta_{\max}(\mathrm{ZA}) = 95\,\mathrm{cm}^{-1}$, $\Delta_{\max}(\mathrm{LO}) = \Delta_{\max}(\mathrm{TO}) = 0.09\,\mathrm{cm}^{-1}$ and $\Delta_{\max}(\mathrm{ZO}) = 1.5\,\mathrm{cm}^{-1}$. The in-plane interactions in BLG are much stronger than the weak van der Waals out-of-plane interactions. For this reason, the deviation of the phonon frequencies in BLG from those in SLG is very small (with exception of $ZA_2$ mode). Similar results were reported for SLG and BLG using the DFT [46], valence force field model of lattice dynamics [12, 29] and the optimized Tersoff and L-J potentials [40]. We note here that although various theoretical approaches can predict different energies for LA, TA and ZA phonons at the *Γ* – point [12-13, 29, 46-49], the descriptions of the phonon mode behavior and the dispersion trends are consistent among the different models. The use of the L-J potential for the description of the inter-layer interaction in FLG usually leads to softening of the low-frequencies modes near the BZ center while the frequencies of all other modes are described accurately.

A comparison of the phonon energy dispersions near $\Gamma$- and $K$-points in BZ of AA-BLG (red dashed curves) and AB-BLG (black curves) is presented in Fig. 4. The frequencies of $LA_2/TA_2$ modes of AA-BLG near BZ center are smaller than in AB-BLG. The maximum deference is at $\Gamma$-point, constituting ~ 4.7 cm$^{-1}$. This energy difference decreases fast with increasing $q$. For $q \geq 1$ nm$^{-1}$ the $LA_2/TA_2$ phonon energies of AA-BLG become indistinguishable from those in AB-BLG. However, the difference in $ZA_2$ mode frequencies remains over the entire BZ and constitutes 3 – 8 cm$^{-1}$. The phonon dispersions of $ZO_1/ZO_2$ and $ZA_1/ZA_2$ modes near $K$-point are linear in AA-BLG and parabolic-like in AB-BLG. It is interesting to note that the evolution of the phonon spectra has similarities to that of the electron spectra near $K$-point in AA-BLG and AB-BLG [43, 44].

The number of atoms in the unit cell of T-BLG with $\theta = 21.8^0$ ($\theta = 13.7^0$) increases by a factor of 7 (19) as compared with BLG. The number of phonon branches increases to 84 for T-BLG with $\theta = 21.8^0$ and to 228 for T-BLG with $\theta = 13.7^0$. The number of phonon modes at $\Gamma$- and $K$-points in BZ of T-BLG increases correspondingly. In addition to the degenerate TO/LO phonon modes of SLG/BLG at $\Gamma$-point with the frequency $\omega$ ~ 1589.5 cm$^{-1}$, the new in-plane phonon modes appear in T-BLG. The frequencies of these modes depend strongly on the rotational angle and their number increases with decreasing $\theta$. In T-BLG with $\theta = 21.8^0$, there appear additional phonon modes at the $\Gamma$-point related to the in-plane optical phonons with the frequencies $\omega$ ~ 1378.6 cm$^{-1}$, 1468.8 cm$^{-1}$ and 1589.5 cm$^{-1}$. In T-BLG with $\theta = 13.7^0$ one can observe new phonon modes with six different frequencies: $\omega$ ~ 1353.0 cm$^{-1}$, 1363.1 cm$^{-1}$, 1367.2 cm$^{-1}$, 1458.8 cm$^{-1}$, 1479.3 cm$^{-1}$ and 1564.1 cm$^{-1}$. Analogously, at $K$-point of BZ instead of the in-plane phonon modes with $\omega$ ~ 1197.3 cm$^{-1}$ and 1347.4 cm$^{-1}$, which are observed in SLG and BLG, one finds the phonon modes with the frequencies $\omega$ ~ 1197.4 cm$^{-1}$, 1347.4 cm$^{-1}$, 1350.7 cm$^{-1}$, 1411.6 cm$^{-1}$, 1486.8 cm$^{-1}$ and 1569 cm$^{-1}$ in T-BLG with $\theta = 21.8^0$. In T-BLG with $\theta = 13.7^0$ the number of different frequencies of $K$-point phonons rises to 14: $\omega$ ~ 1197.4 cm$^{-1}$, 1260.5 cm$^{-1}$, 1339.8 cm$^{-1}$, 1347.4 cm$^{-1}$, 1351.4 cm$^{-1}$, 1365.2 cm$^{-1}$, 1390.4 cm$^{-1}$, 1395.6 cm$^{-1}$, 1449.6 cm$^{-1}$, 1491.5 cm$^{-1}$, 1498.2 cm$^{-1}$, 1547.3 cm$^{-1}$, 1552.8 cm$^{-1}$ and 1584.7 cm$^{-1}$. Overall the phonon spectrum of twisted bilayer graphene becomes much more complicated.

The twisting influences the phonon spectra of BLG owing to two reasons: (i) modification of the weak van der Waals inter-layer interaction and (ii) alteration of a size of a BZ leading to the phonon momentum change. To investigate these effects separately in Fig. 5 (a, b) we plot the phonon dispersions in AA-BLG along $\Gamma_0$-$K_0$ (red curves) and $\Gamma_1$-$K_1$ (blue curves) direction in BZ of AA-BLG BZ. In BZ of AA-BLG there are seven directions $\Gamma_i$-$K_i$ (i = 0,...,6) which are equivalent to the direction $\Gamma_0$-$K_0$ in BZ of T-BLG BZ (see Fig. 5 (c), where the Brillouin zones for the BLG and T-BLG are shown). The phonon dispersion in T-BLG with $\theta = 21.8^0$, shown in Fig. 3 (c), corresponds to the phonon dispersions along $\Gamma_i$-$K_i$ (i = 0,...,6) directions in BZ of AA-BLG. Thus, in the twisted bilayer graphene appear *entangled* phonon branches resulting from mixing of different directions from BLG BZ: $\Gamma_0$-$K_0$, $\Gamma_1$-$K_1$, .., $\Gamma_6$-$K_6$. The red and blue curves in Fig. 3 (c) appear in the phonon spectra of T-BLG from $\Gamma_0$-$K_0$ and $\Gamma_1$-$K_1$ direction in BZ of AA-BLG. The difference in the frequencies of the phonon modes shown in Figs. 3 (c) (red and blue curves) and 5 (a-b) is a manifestation of the twisting. In T-BLG the difference between the phonon frequencies of all corresponding modes is small due to weak inter-layer interaction. However, this difference may be higher for other layered *van der Waals* materials with stronger interlayer coupling.

**<Table IV>**

The phonon frequencies of the in-plane optical phonons at *Γ*- and *K*-points are presented in Table IV for SLG, BLG and T-BLG with the different rotation angles $\theta(1,1) = 21.8^0$, $\theta(2,1) = 13.2^0$, $\theta(3,1) = 9.4^0$ and $\theta(4,1) = 7.3^0$. In the round parenthesis we show the number of phonon modes in the indicated energy range, i.e. 1350.7-1350.8 (6) means that there are 6 near-degenerate phonon modes in the energy interval 1350.7 – 1350.8 $cm^{-1}$. The difference between the frequencies of corresponding TO/LO modes in SLG/BLG and T-BLG is small, constituting less than 0.1 $cm^{-1}$ at *K*-point and less than 0.05 $cm^{-1}$ at *Γ*-point. The latter means that the rotationally-dependent modification of the interlayer interaction changes frequencies of in-plane optical modes by less than 0.1 $cm^{-1}$ over the entire BZ. In Table V, we present data for *Γ*-point frequencies of ZO modes in T-BLG. Analogously to TO/LO phonons this mode is only weakly dependent on twisting: the maximum frequency difference is ~0.5 $cm^{-1}$.

**<Table V>**

The frequencies of the shear ($LA_2$, $TA_2$) and flexural ($ZA_2$) phonons are affected stronger. The specific properties of these modes in T-BLG with $\theta = 21.8^0$ (red curves) and T-BLG with $\theta = 13.7^0$ (blue curves) as well as in AA-BLG (black curves) are presented in Fig. 6. At $\Gamma$ – point, the twisting increases the frequency of the shear modes by 1 – 2 cm$^{-1}$ and decreases the frequency of $ZA_2$ modes by ~ 5 – 5.5 cm$^{-1}$ depending on $\theta$. The detailed dependence of the acoustic phonon frequencies on $\theta$ is presented in Table VI.

**<Table VI>**

## IV. Discussion and Comparison with Experiments

Our results are in agreement with an intuitive expectation that changes in the weak van der Waals interaction should not affect the phonon modes significantly. Nevertheless, the new entangled phonon modes at $\Gamma$-, $K$- and $M$-points in BZ of T-BLG can be observed in Raman or infrared spectra and possibly used for determination of the rotation angle in T-BLG samples. Most recently the Raman peak near 1489 cm$^{-1}$ was observed in CVD-synthesized T-BLG with $\theta = 13.2^0$ [50], which was attributed to the folded phonons. Using a simple model for folding of the optical phonon dispersion in SLG, the authors estimated the frequency of this peak as $\omega$ = 1480 cm$^{-1}$ [50]. Our calculations show that in T-BLG with $\theta = 13.2^0$ at $K$-point there are entangled phonon modes with $\omega$ = 1491.5 cm$^{-1}$, which is closer to experimental value. A conclusive assessment of the nature of the observed peak requires further experimental studies and comparison with the calculated dispersion.

The final Moire pattern in T-BLG depends both on the (i) initial stacking configuration and (ii) location of the rotation origin. For a proper comparison of the results obtained within the approach where the initial stacking is AB with the results obtained within our rotation scheme it is essential that the axis of the rotation from AB-stacking passes through the atoms, which lie exactly above each other. This ensures that for the rotation angles between 0 and 30 degrees the Moire patterns will coincide and the results will be directly comparable.

Our results – phonon dispersion and tabulated frequencies – can be used for interpretation of experimental Raman peaks reported by different groups for T-BLG samples even when no accurate measurements of $\theta$ were carried out [51, 52]. A comparison between the theoretical and experimental data and our assessment of the rotational angles are presented in Table VII. The frequencies of the entangled phonons at *Γ*-, *M*-, and *K*-points calculated in this work for $\theta$=21.8$^0$ 13.2$^0$, 9.4$^0$ and 7.3$^0$ are matching closely the measured Raman frequencies for most of the considered samples [51, 52]. Additional calculations with different rotational schemes and various angles are necessary for interpretation of the peaks frequencies for S2, S4, S5 and S8 samples from Ref. [52]. This work is reserved for the future study.

The Raman measurements with T-BLG with $\theta \sim 9^0 - 16^0$ demonstrated a set of peaks in the frequency ranges 100 – 900 cm$^{-1}$ and 1400 – 1600 cm$^{-1}$ [23]. The authors compared their results with the phonon dispersions in SLG and concluded that these peaks are related to ZA, TA, LA, ZO, TO and LO phonon branches [23]. Our results, in general, confirm their conclusion with one important refinement. The nature of these peaks is more complicated since they are made up of the in-plane and out-of-plane entangled phonons from the rotationally-dependent BZ of T-BLG. For instance, the Raman peaks with the frequencies 100 – 200 cm$^{-1}$ can be associated not only with $ZA_2$ (ZO') branch as concluded in Ref. [23] but with the entangled $ZA_1$ and $ZA_2$ branches (see Table VI). The peaks with the frequencies of 300 – 900 cm$^{-1}$ can be related to the entangled $ZA_1/ZA_2$, $TA_1/TA_2$, $LA_1/LA_2$ and $ZO_1/ZO_2$ phonons while the peaks with the frequencies 1400 – 1600 cm$^{-1}$ are associated with the entangled $TA_1/TA_2$ and $LA_1/LA_2$ branches (see Tables V-VII). Complete understanding of the entangled phonon modes contribution to Raman processes in T-BLGs requires further closely correlated computational and experimental studies with accurate measurement of the rotational angle in T-BLG.

**<Table VII>**

The two-dimensional (2D) phonon density of states $f(\omega) = \sum_{s(\omega)} (d\omega_s / dq_y)^{-1}$ and phonon average group velocity $\langle \upsilon \rangle(\omega) = N(\omega) / \sum_{s(\omega)} (d\omega_s / dq_y)^{-1}$ as the functions of the phonon frequency are presented in Fig. 7 and 8, respectively. The data are shown for the AA-BLG (black curves) and T-BLG with $\theta = 21.8^0$ (red curves) and $\theta = 13.7^0$ (blue curves). The summation is performed over all phonon modes $s$ with frequency $\omega$ from the entire BZ of BLG and BZ of T-BLG BZ. In the above definition $N(\omega)$ is the number of phonon modes with the frequency $\omega$. As follows from Figs. 7 and 8, the changes in the phonon DOS and average group velocity due to twisting are small. The difference in the phonon DOS appear only in one frequency range of $90-110\,\mathrm{cm}^{-1}$. The effect is due to the frequency shift of $ZA_2$ phonons. Therefore we do not expect a major difference in the phonon-assisted processes in T-BLG in comparison with BLG as a result of changes in the phonon DOS or average group velocity. However, the rotationally-dependent entangled phonons with the momenta different from those of the corresponding phonon modes in BLG can significantly change the electron-phonon and phonon-phonon processes in T-BLG. The later can lead to the rotationally-dependent electron-phonon interaction and thermal conductivity.

## V. Conclusions

We have theoretically investigated the phonon properties of AA-stacked, AB-stacked and twisted bilayer graphene, using the Born-von-Karman model for the intra-layer interactions and Lennard-Jones potential for the inter-layer interactions. It was established that $ZA_2$ phonon mode is affected by the stacking order the most. The frequency of $ZA_2$ in AA-BLG is smaller than in AB-BLG by 3 – 8 $cm^{-1}$ depending on the phonon wave vector. Twisting in bilayer graphene leads to an additional decreases in the frequency of this mode by 5 – 5.5 $cm^{-1}$ depending on the rotation angle. We have shown that a new type of *entangled* rotationally-dependent phonon modes appear in the twisted bilayer graphene due to reduction of the Brillouin zone size and changes in the high-symmetry directions. These modes can manifest themselves in Raman or infrared measurements and, thus, can be used for the non-contact characterization of twisted bilayer or multilayer graphene. It was also demonstrated that twisting-induced change in the van der Waals inter-layer interactions weakly affects the phonon frequencies. The changes in phonon density of states and average group velocities in T-BLG as compared to those in AA- or AB-stacked bilayer graphene are small. The obtained results and tabulated frequencies of phonons in twisted bilayer graphene are important for interpretation of the experimentally measured Raman and infrared spectra and understanding thermal properties of these systems.

***Acknowledgements***

The work at UCR was supported by the Semiconductor Research Corporation (SRC) and the Defense Advanced Research Project Agency (DARPA) through FCRP Center for Function Accelerated nanoMaterial Engineering (FAME) and by the National Science Foundation (NSF) projects EECS-1128304, EECS-1124733 and EECS-1102074. DLN and AIC acknowledge the financial support from the Moldova State projects no. 11.817.05 and the National Scholarship of the World Federation of Scientists (WFS). AAB acknowledges useful discussions of twisted bilayer graphene with Prof. A.C. Ferrari (Cambridge), Prof. P. Kim (Columbia) and Prof. R. Lake (UC Riverside), which stimulated the initial interest to this material system.

**Figure Captions**

**Figure 1**: Schematic of the lattice structure (a) and Brillouin zone (b) in the twisted bilayer graphene with the rotational angle $\theta = 21.8^0$.

**Figure 2:** Four coordination spheres in the single layer graphene used in the theoretical description.

**Figure 3**: Phonon energy dispersions in single layer graphene (a), AA-stacked bilayer graphene (b), and in the twisted bilayer graphene with $\theta = 21.8^0$ (c) and $\theta = 13.2^0$ (d). All dispersion relations are shown for $\Gamma - K$ direction in the Brillouin zone of SLG, BLG and T-BLG, correspondingly.

**Figure 4:** Phonon energy dispersions near $\Gamma$- point (a) and near $K$-point (b) of the Brillouin zone in AA-stacked (red dashed curves) and AB-stacked (black solid curves) bilayer graphene.

**Figure 5:** Phonon energy dispersions in AA-stacked bilayer graphene shown for $\Gamma_0$-$K_0$ (a) and $\Gamma_1$-$K_1$ (b) directions of the Brillouin zone of AA-BLG. (c) Brillouin zones in AA-BLG and T-BLG with $\theta = 21.8^0$. Note that seven high-symmetry directions $\Gamma_i$-$K_i$ (i = 0,..., 6) in the Brillouin zone of AA-BLG are equivalent to the high-symmetry direction $\Gamma_0$-$K_0$ in the Brillouin zone of T-BLG.

**Figure 6:** Zone-center phonon dispersions of the out-of-plane (a) and in-plane (b) acoustic modes in AA-BLG (black curves), T-BLG with $\theta = 21.8^0$ (blue curves) and T-BLG with $\theta = 13.2^0$ (red curves). Note how a change in the interlayer interaction due to twisting affects $TA_2$, $LA_2$, and $ZA_2$ modes.

**Figure 7:** Two-dimensional phonon density of states as a function of the phonon frequency in AA-BLG (black curves), T-BLG with $\theta = 21.8^0$ (blue curves) and T-BLG with $\theta = 13.2^0$ (red curves). Phonon DOS are presented for an extended frequency range (a) and for the frequency interval of $90\,\mathrm{cm}^{-1} - 110\,\mathrm{cm}^{-1}$ where the maximum difference in the phonon DOS is observed.

**Figure 8**: Phonon average group velocity as a function of the phonon frequency in AA-BLG (black curves), T-BLG with $\theta = 21.8^0$ (blue curves) and T-BLG with $\theta = 13.2^0$ (red curves).

**Table I: Parameters if the theoretical model**

| Atom # (*j*) | I[st] sphere (*s*=1) | II[nd] sphere (*s*=2) | III[rd] sphere (*s*=3) | IV[th] sphere (*s*=4) |
|---|---|---|---|---|
| 1 | $(a,0)$ | $(0,\sqrt{3}a)$ | $(-2a,0)$ | $(5a/2,\sqrt{3}a/2)$ |
| 2 | $(-a/2,-\sqrt{3}a/2)$ | $(3a/2,\sqrt{3}a/2)$ | $(a,\sqrt{3}a)$ | $(5a/2,-\sqrt{3}a/2)$ |
| 3 | $(-a/2,\sqrt{3}a/2)$ | $(3a/2,-\sqrt{3}a/2)$ | $(a,-\sqrt{3}a)$ | $(-a/2,-3\sqrt{3}a/2)$ |
| 4 | – | $(0,-\sqrt{3}a)$ | – | $(-2a,-\sqrt{3}a)$ |
| 5 | – | $(-3a/2,-\sqrt{3}a/2)$ | – | $(-2a,\sqrt{3}a)$ |
| 6 | – | $(-3a/2,\sqrt{3}a/2)$ | – | $(-a/2,3\sqrt{3}a/2)$ |

**Table II: Force constant matrices**

| **Atom # (*j*)** | **I^st sphere (*s*=1)** | **II^nd sphere (*s*=2)** | **III^rd sphere (*s*=3)** | **IV^th sphere (*s*=4)** |
|---|---|---|---|---|
| 1 | $\Phi^{(I)}$ | $\Phi^{(II)}$ | $\Phi^{(III)}$ | $\sigma_y\left(R_{\pi/6}\Phi^{(IV)}R_{\pi/6}^{-1}\right)\sigma_y^{-1}$ |
| 2 | $R_{2\pi/3}\Phi^{(I)}R_{2\pi/3}^{-1}$ | $\sigma_y\left(R_{2\pi/3}\Phi^{(II)}R_{2\pi/3}^{-1}\right)\sigma_y^{-1}$ | $R_{2\pi/3}\Phi^{(III)}R_{2\pi/3}^{-1}$ | $R_{\pi/6}\Phi^{(IV)}R_{\pi/6}^{-1}$ |
| 3 | $R_{4\pi/3}\Phi^{(I)}R_{4\pi/3}^{-1}$ | $R_{2\pi/3}\Phi^{(II)}R_{2\pi/3}^{-1}$ | $R_{4\pi/3}\Phi^{(III)}R_{4\pi/3}^{-1}$ | $\sigma_y\left(R_{4\pi/3}\Phi^{(IV)}R_{4\pi/3}^{-1}\right)\sigma_y^{-1}$ |
| 4 | – | $\sigma_y\Phi^{(II)}\sigma_y^{-1}$ | – | $R_{2\pi/3}\Phi^{(IV)}R_{2\pi/3}^{-1}$ |
| 5 | – | $R_{4\pi/3}\Phi^{(II)}R_{4\pi/3}^{-1}$ | – | $\sigma_y\left(R_{2\pi/3}\Phi^{(IV)}R_{2\pi/3}^{-1}\right)\sigma_y^{-1}$ |
| 6 | – | $\sigma_y\left(R_{4\pi/3}\Phi^{(II)}R_{4\pi/3}^{-1}\right)\sigma_y^{-1}$ | – | $R_{4\pi/3}\Phi^{(IV)}R_{4\pi/3}^{-1}$ |

**Table III: Intra-layer interatomic force constants for graphene**

| **Force constants** | **I$^{st}$ sphere ($s$=1)** | **II$^{nd}$ sphere ($s$=2)** | **III$^{rd}$ sphere ($s$=3)** | **IV$^{th}$ sphere ($s$=4)** |
|---|---|---|---|---|
| $\alpha$ (N/m) | 398.7 | 72.9 | -26.4 | 1.0 |
| $\beta$ (N/m) | 172.8 | -46.1 | 33.1 | 7.9 |
| $\gamma$ (N/m) | 98.9 | -8.2 | 5.8 | -5.2 |

**Table IV: Frequencies of the in-plane optical phonons at *Γ*- and *K*-points ($cm^{-1}$)**

| SLG | BLG | T-BLG $\theta$=21.8$^0$ | T-BLG $\theta$=13.2$^0$ | T-BLG $\theta$=9.4$^0$ | T-BLG $\theta$=7.3$^0$ |
|---|---|---|---|---|---|
| *Γ*-point | | | | | |
| 1589.45 (2) | 1589.43 (2)<br>1589.52 (2) | 1378.6-1378.7 (12)<br>1468.8-1468.9 (12)<br>1589.47 (4) | 1353.0-1353.1 (12)<br>1363.1-1363.2 (12)<br>1367.2-1367.3 (12)<br>1458.8 (12)<br>1479.3-1479.4 (12)<br>1564.1-1564.2 (12)<br>1589.48 (4) | 1312.9-1313.0 (12)<br>1348.8-1350.0 (24)<br>1354.2 (12)<br>1407.9-1408.3 (24)<br>1417.9 (12)<br>1434.0 (12)<br>1509.5 (12)<br>1527.1 (12)<br>1543.8 (12)<br>1580.2-1580.3 (12)<br>1589.48 (4) | 1213.8-1214.7 (24)<br>1282.5-1282.6 (12)<br>1345.6-1345.8 (24)<br>1348.6-1350.1 (24)<br>1367.5 (12)<br>1377.6-1378.7 (24)<br>1395.8-1395.9 (12)<br>1398.5 (12)<br>1409.4 (12)<br>1458.5-1459.4 (24)<br>1483.5 (12)<br>1506.4 (12)<br>1509.8 (12)<br>1550.1 (12)<br>1553.4 (12)<br>1568.5 (12)<br>1584.9-1585.0 (12)<br>1589.48 (4) |
| *K*-point | | | | | |
| 1197.3<br>1347.4 | 1197.3 (2)<br>1347.4<br>1347.5 | 1197.4 (2)<br>1347.4 (2)<br>1350.7-1350.8 (6)<br>1411.6-1411.7 (6)<br>1486.8 (6)<br>1569.0-1569.1 (6) | 1197.4 (2)<br>1260.5 (6)<br>1339.8-1340.8 (12)<br>1347.4 (2)<br>1351.4 (6)<br>1365.2 (6)<br>1390.4 (6)<br>1395.6 (6)<br>1449.6 (6)<br>1491.5 (6)<br>1498.2 (6)<br>1547.3 (6)<br>1552.8 (6)<br>1584.7 (6) | 1197.4 (2)<br>1259.6 (6)<br>1262.9 (6)<br>1299.7 (6)<br>1317.4 (6)<br>1340.2 (6)<br>1344.0 (6)<br>1347.4 (2)<br>1348.7 (6)<br>1363.2 (6)<br>1367.4 (6)<br>1371.1-1371.9 (12)<br>1394.2 (6)<br>1398.0 (6)<br>1424.0 (6)<br>1430.3 (6)<br>1461.5-1462.9 (12)<br>1476.7 (6)<br>1501.5 (6)<br>1506.4-1508.2 (12)<br>1555.8-1557.3 (12)<br>1567.2 (6)<br>1576.4 (6)<br>1587.5 (6) | 1197.4 (2)<br>1251.8 (6)<br>1254.7 (6)<br>1274.4-1276.8 (12)<br>1284.7 (6)<br>1339.0-1341.0 (12)<br>1345.6 (6)<br>1347.4 (2)<br>1348.2-1350.1 (12)<br>1354.3 (6)<br>1356.6-1357.4 (12)<br>1360.1-1362.4 (12)<br>1370.4 (6)<br>1374.4 (6)<br>1376.3-1376.6 (12)<br>1381.6 (6)<br>1411.0-1412.7 (12)<br>1425.6 (6)<br>1438.9 (6)<br>1444.7 (6)<br>1448.9 (6)<br>1450.3 (6)<br>1452.6 (6)<br>1458.4 (6)<br>1486.1 (6)<br>1504.4-1504.9 (12)<br>1514.2-1516.3 (12)<br>1537.6-1538.2 (12)<br>1549.9-1551.6 (12)<br>1574.1-1575.8 (18)<br>1583.1 (6)<br>1588.4 (6) |

**Table V: Frequencies of the out-of-plane optical modes at $\Gamma$-point ($cm^{-1}$)**

| **SLG** | **BLG** | **T-BLG $\theta$=21.8$^0$** | **T-BLG $\theta$=13.2$^0$** | **T-BLG $\theta$=9.4$^0$** | **T-BLG $\theta$=7.3$^0$** |
|---|---|---|---|---|---|
| 894.1 | 895.9<br>897.4 | 725.6-728.2 (12)<br>896.36 (2) | 651.3-653.1 (12)<br>707.4-708.0 (12)<br>836.8-837.7 (12)<br>896.34 (2) | 619.6-621.8 (12)<br>672.9-675.5 (24)<br>769.4-769.5 (12)<br>802.1 (12)<br>866.7-867.5 (12)<br>896.34 (2) | 602.6 (6)<br>604.6-605.0 (6)<br>650.1-652.9 (24)<br>666.7-667.6 (12)<br>721.8 (12)<br>759.3-760.7 (24)<br>821.6-821.7 (12)<br>841.2 (12)<br>878.6-879.4 (12)<br>896.35 (2) |

**Table VI: Frequencies of the in-plane optical phonons at $\Gamma$-point (in $cm^{-1}$)**

| **SLG** | **BLG** | **T-BLG $\theta$=21.8$^0$** | **T-BLG $\theta$=13.2$^0$** | **T-BLG $\theta$=9.4$^0$** | **T-BLG $\theta$=7.3$^0$** |
|---|---|---|---|---|---|
| in-plane | | | | | |
| 0 (2) | 0 (2)<br>13.4 (2) | 0 (2)<br>14.7 (2)<br>688.1-688.4 (12)<br>1025.3-1025.4 (12) | 0 (2)<br>13.8 (2)<br>465.6-465.9 (12)<br>602.2-602.4 (12)<br>682.2-682.4 (12)<br>840.9-841 (12)<br>1118.5-1118.6 (12)<br>1128.3-1128.4 (12) | 0 (2)<br>13.8 (2)<br>341.2-341.5 (12)<br>500.4-500.6 (12)<br>505.0-505.3 (12)<br>618.2-618.2 (12)<br>650.0-650.2 (12)<br>714.6-714.7 (12)<br>855.2-855.5 (12)<br>905.9-906 (12)<br>908.7-908.9 (12)<br>1149.5-1149.6 (12)<br>1155.4-1155.5 (12)<br>1193.5-1193.6 (12) | 0 (2)<br>14.0 (2)<br>267.9-268.3 (12)<br>399.0-399.4 (12)<br>417.9-418.2 (12)<br>520.0-520.3 (12)<br>562.0-562.1 (12)<br>603.4-603.5 (12)<br>606.9-607.1 (12)<br>682.9-683.2 (12)<br>694.11-694.12 (12)<br>745.4-745.6 (12)<br>749.6-749.8 (12)<br>796.63-796.65 (12)<br>945.2-945.3 (12)<br>964.63-964.65 (12)<br>986.61-986.63 (12)<br>1002.7-1002.8 (12)<br>1161.93-1161.95 (12)<br>1166.55-1166.62 (12) |
| out-of-plane | | | | | |
| 0 | 0<br>95.0 | 0<br>89.5<br>336.8-337.9 (6)<br>345.6-347.7 (6) | 0<br>89.3<br>147.4-147.9 (6)<br>169.9-172 (6)<br>369.1-369.2 (6)<br>376.9-377.4 (6)<br>433.0-433.2 (6)<br>438.9-439.6 (6) | 0<br>79.9-80.4 (6)<br>89.2<br>117.6-120.4 (6)<br>218.48-218.52 (6)<br>234.0-234.1 (6)<br>271.2-271.3 (6)<br>283.31-283.33 (6)<br>411.2-411.6 (12)<br>418.3-420.2 (12)<br>468.6-468.8 (6)<br>473.6-474.2 (6) | 0<br>49.86-50.28 (6)<br>89.1<br>100.2-103.5 (6)<br>140.4-140.5 (6)<br>164.7-164.9 (6)<br>179.01-179.06 (6)<br>198.2-198.23 (6)<br>289.0-289.3 (12)<br>300.1-301.2 (12)<br>342.54-342.56 (6)<br>351.36-351.37 (6)<br>423.9-430.2 (12)<br>437.7-437.8 (12)<br>443.6-443.9 (12)<br>486.9-487.1 (6)<br>491.5-492 (6) |

**Table VII: Theoretical and experimental phonon frequencies in T-BLG**

| Reported Data* | Sample | Raman peak frequency from experiment ($cm^{-1}$) | Entangled phonon frequency from the theory ($cm^{-1}$) | Assumed rotational angle |
|---|---|---|---|---|
| A | S1, S2, S4 | 1351,1352, 1352<br>1384,1382,1384 | 1350.1 (*K*-point), 1351.5 (*M*-point)<br>1381.6 (*K*-point) | $\Theta$~7.3$^0$ |
| | S3 | 1352<br>1371 | 1350 (*Γ*-point), 1354.6 (*M*-point)<br>1371.1 (*K*-point) | $\Theta$~9.4$^0$ |
| | S5 | 1355<br>1395 | 1353 (*Γ*-point), 1357.5 (*M*-point)<br>1395.6 (*K*-point) | $\Theta$~13.2$^0$ |
| B | S1 | 1110<br>1391<br>1452<br>1588 | 1109, 1111 (*M*-point)<br>1389,1391 (*M*-point)<br>1452.6 (*K*-point)<br>1588.4 (*K*-point)<br>1588.6 (*M*-point) | $\Theta$~7.3$^0$ |
| | S3 | 1134<br>1392<br><br>1443 | 1131 (*K*-point)<br>1392 (*M*-point)<br>1395 (*Γ*-point)<br>1444 (*K*-point) | $\Theta$~7.3$^0$ |
| | S6 | 1383<br>1424<br>1588 | 1382.6 (*M*-point )<br>1424 (*K*-point )<br>1588 (*M*-point) | $\Theta$~9.4 |
| | S7 | 1379<br><br>1395<br>1434<br>1595 | 1377.4 (*M*-point)<br>1381.4 (*M*-point)<br>1394.2 (*K*-point )<br>1434 (*Γ*-point)<br>1589.5 (*Γ*-point) | $\Theta$~9.4 |

*Experimental data set A is from Ref. [51] while data set B is from Ref. [52]. All theoretical values are from this work.

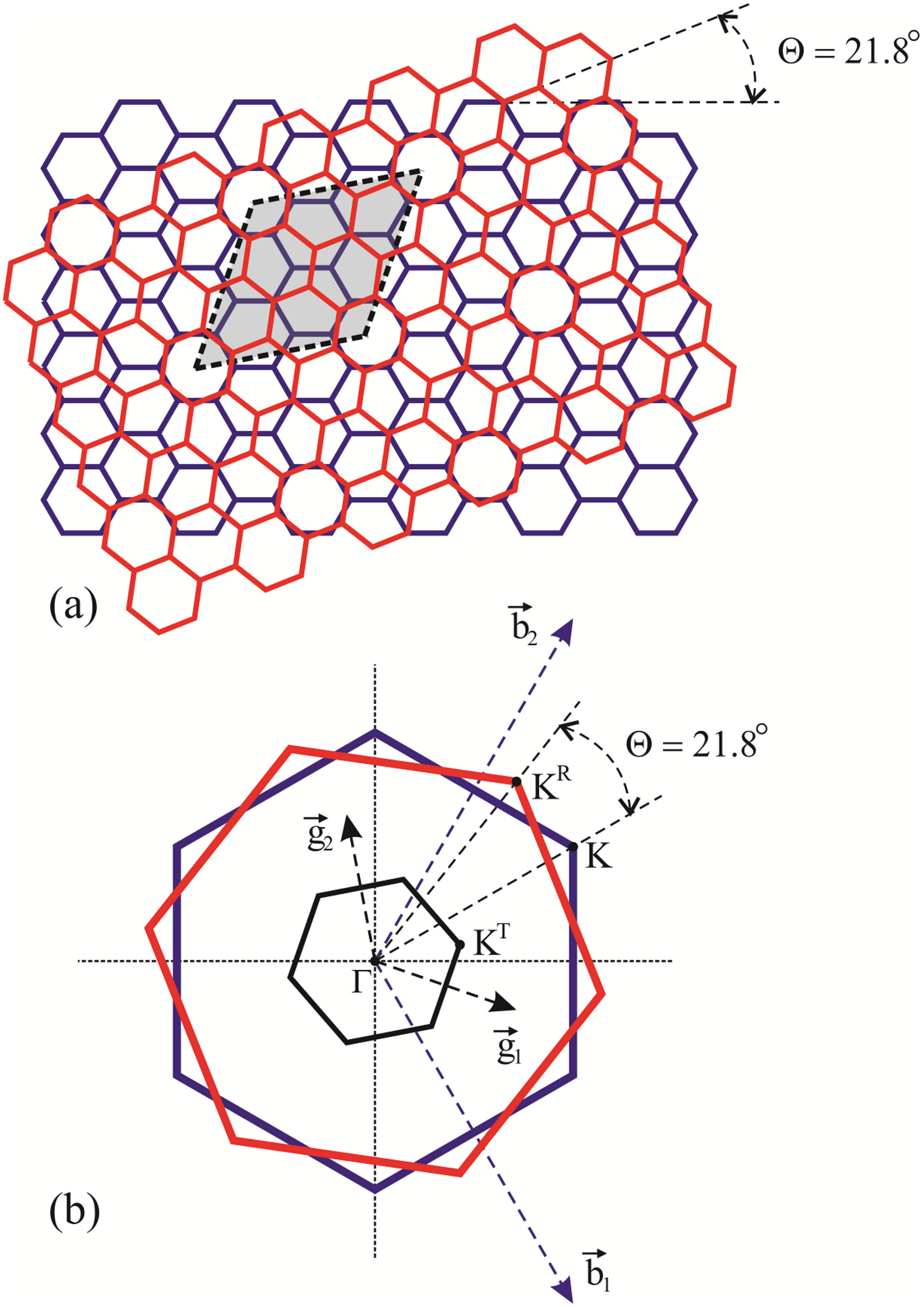

Figure 1

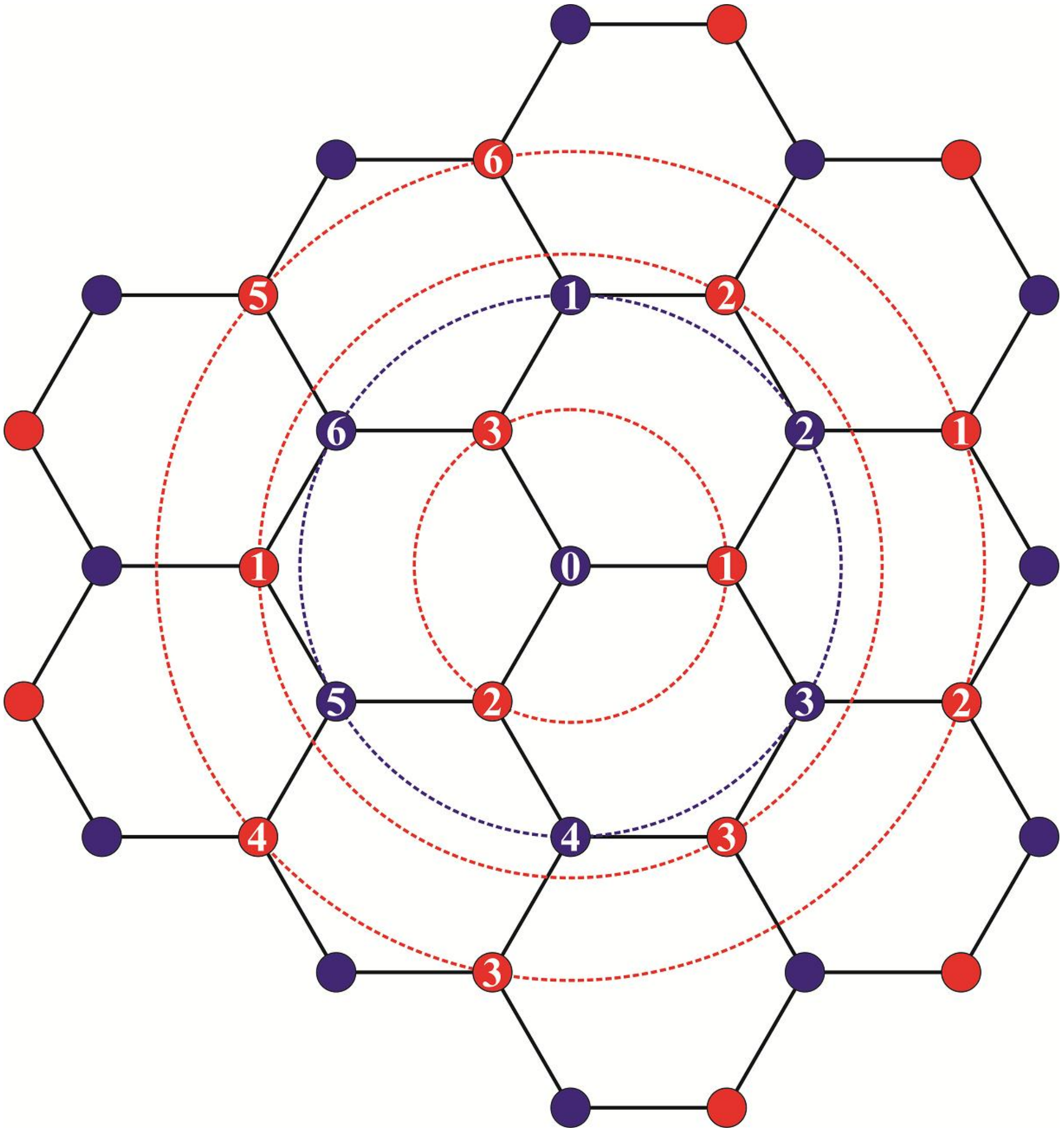

Figure 2

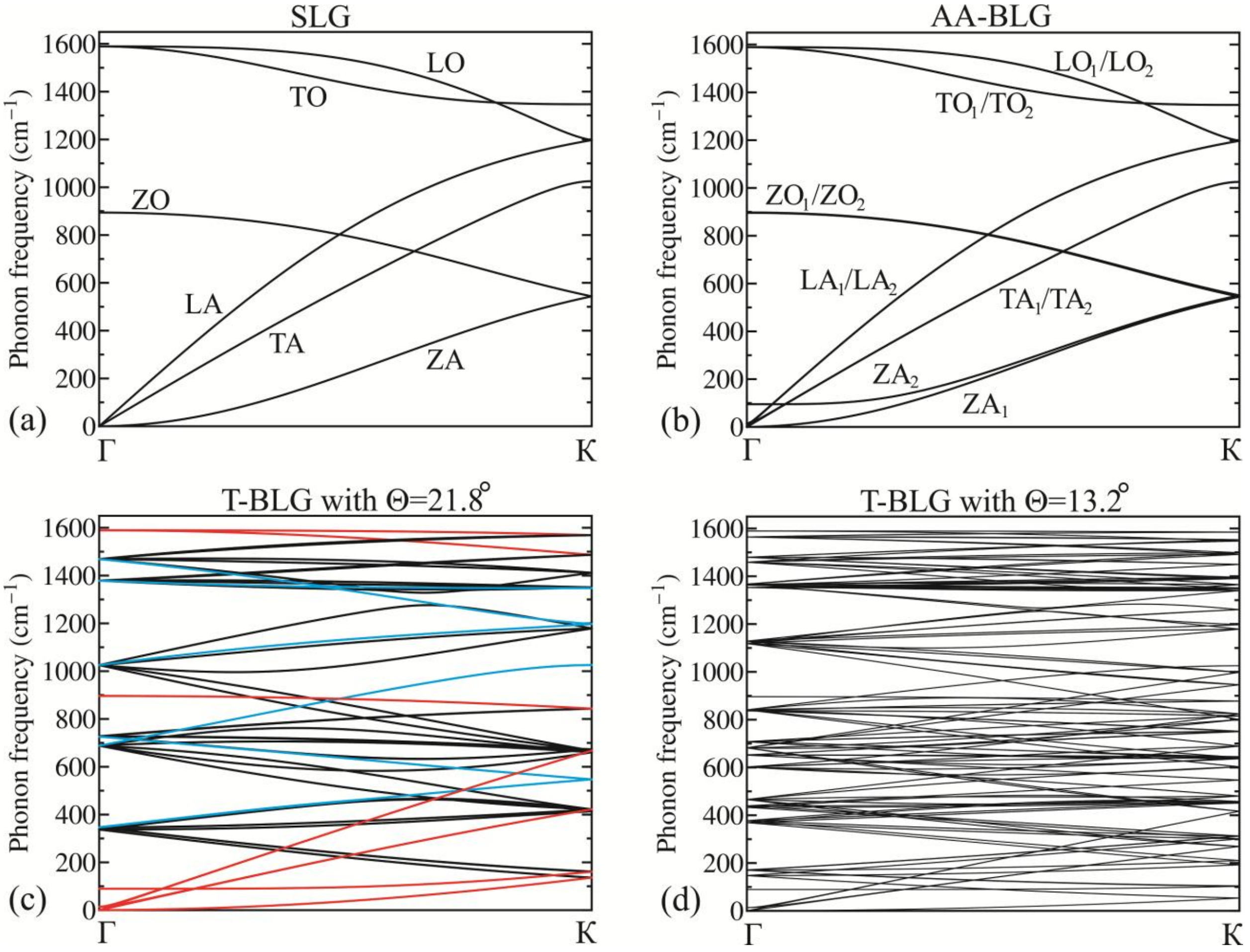

Figure 3

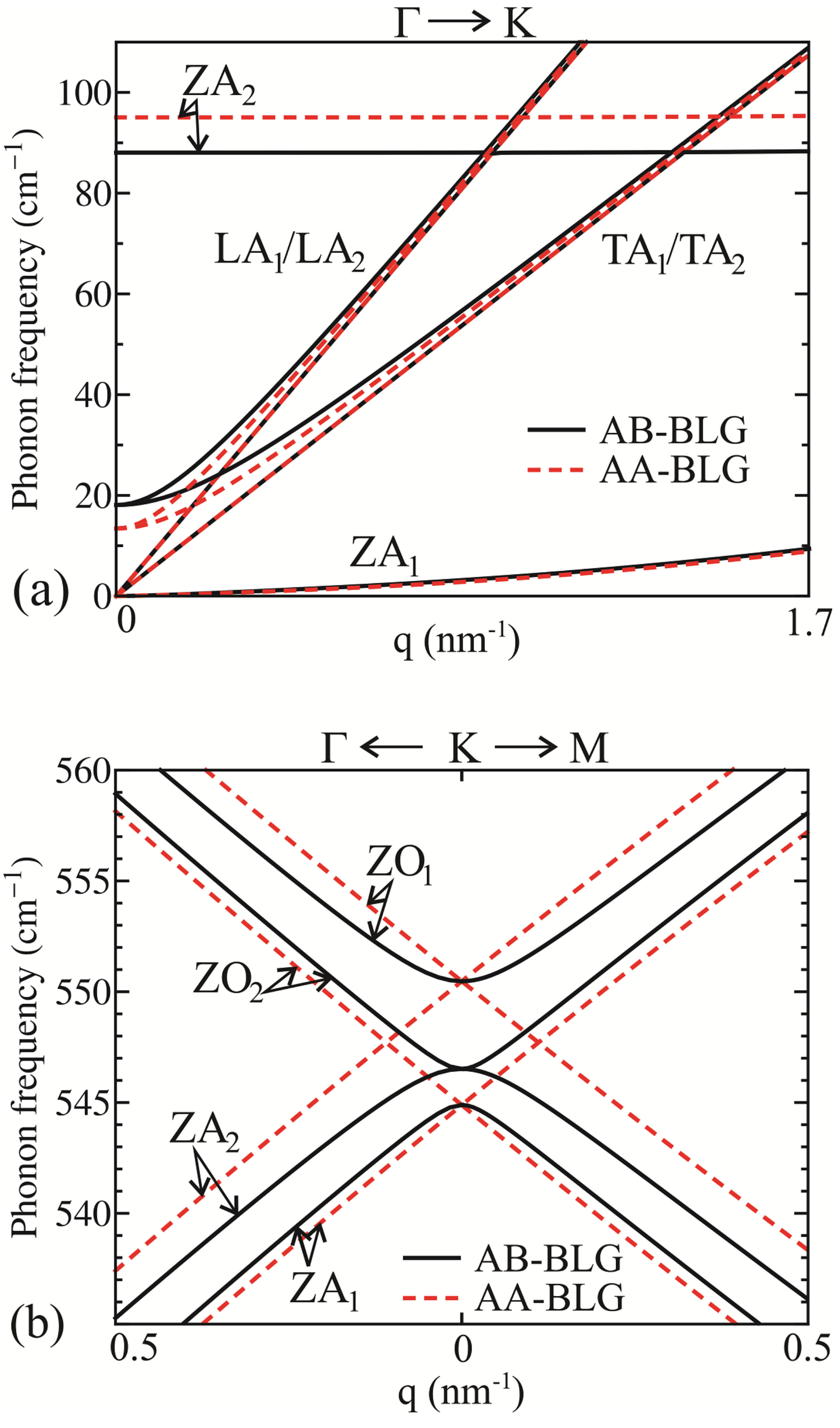

Figure 4

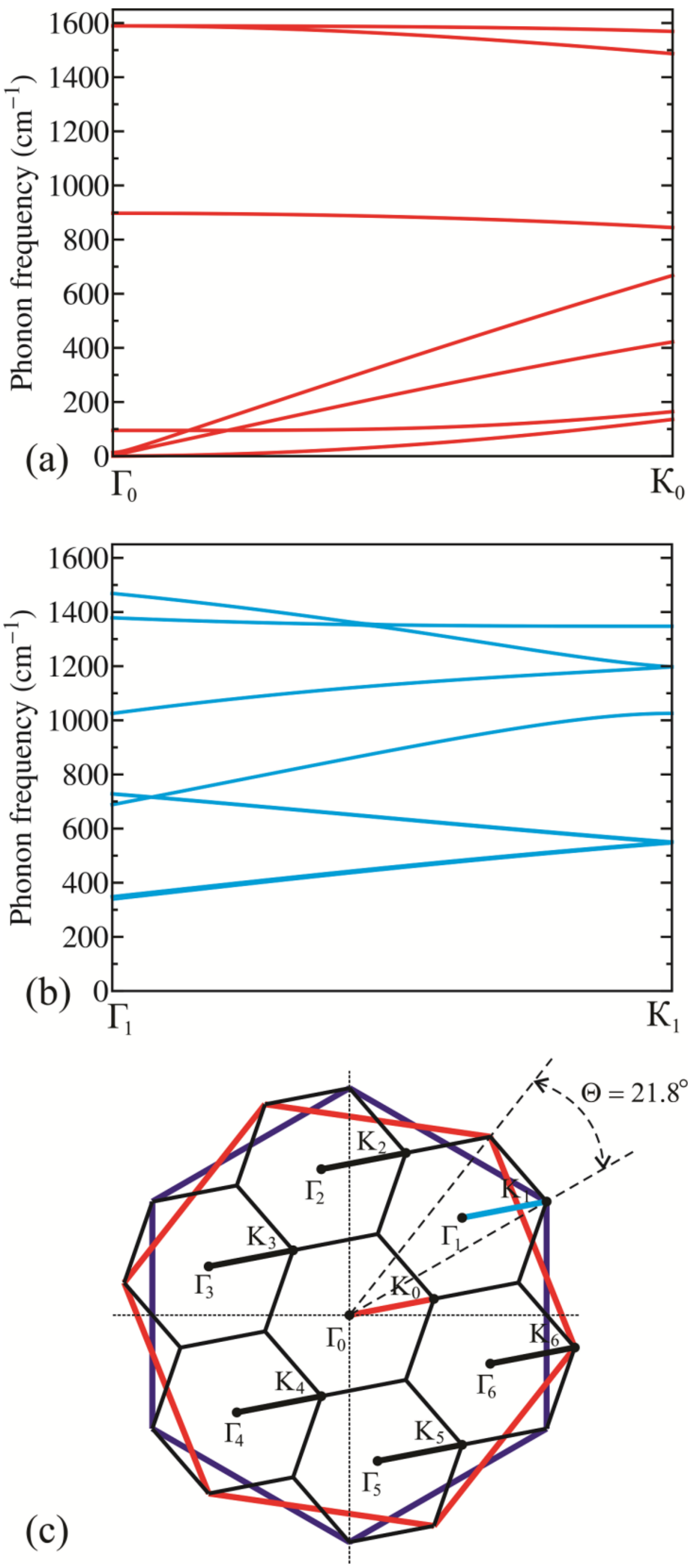

Figure 5

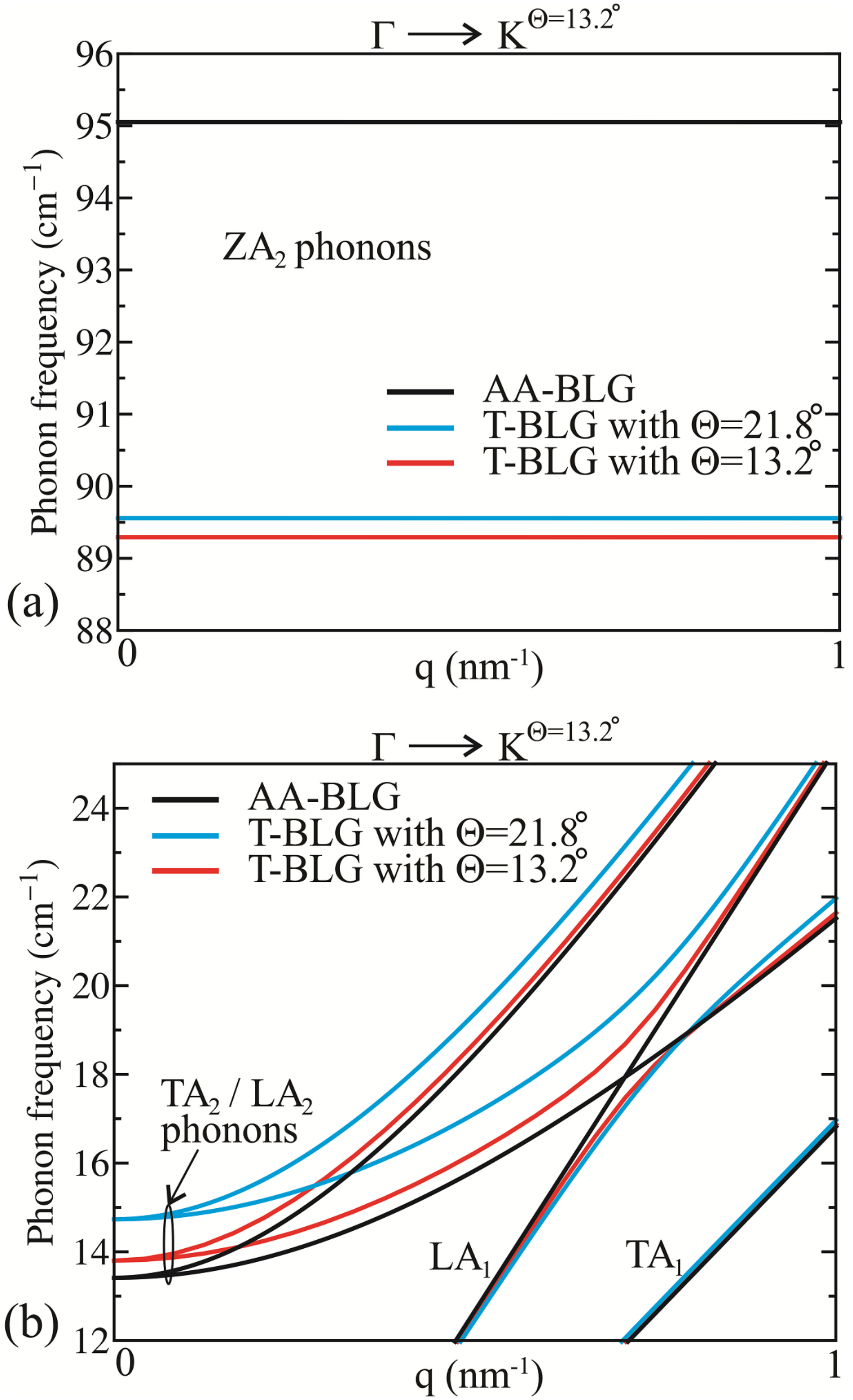

Figure 6

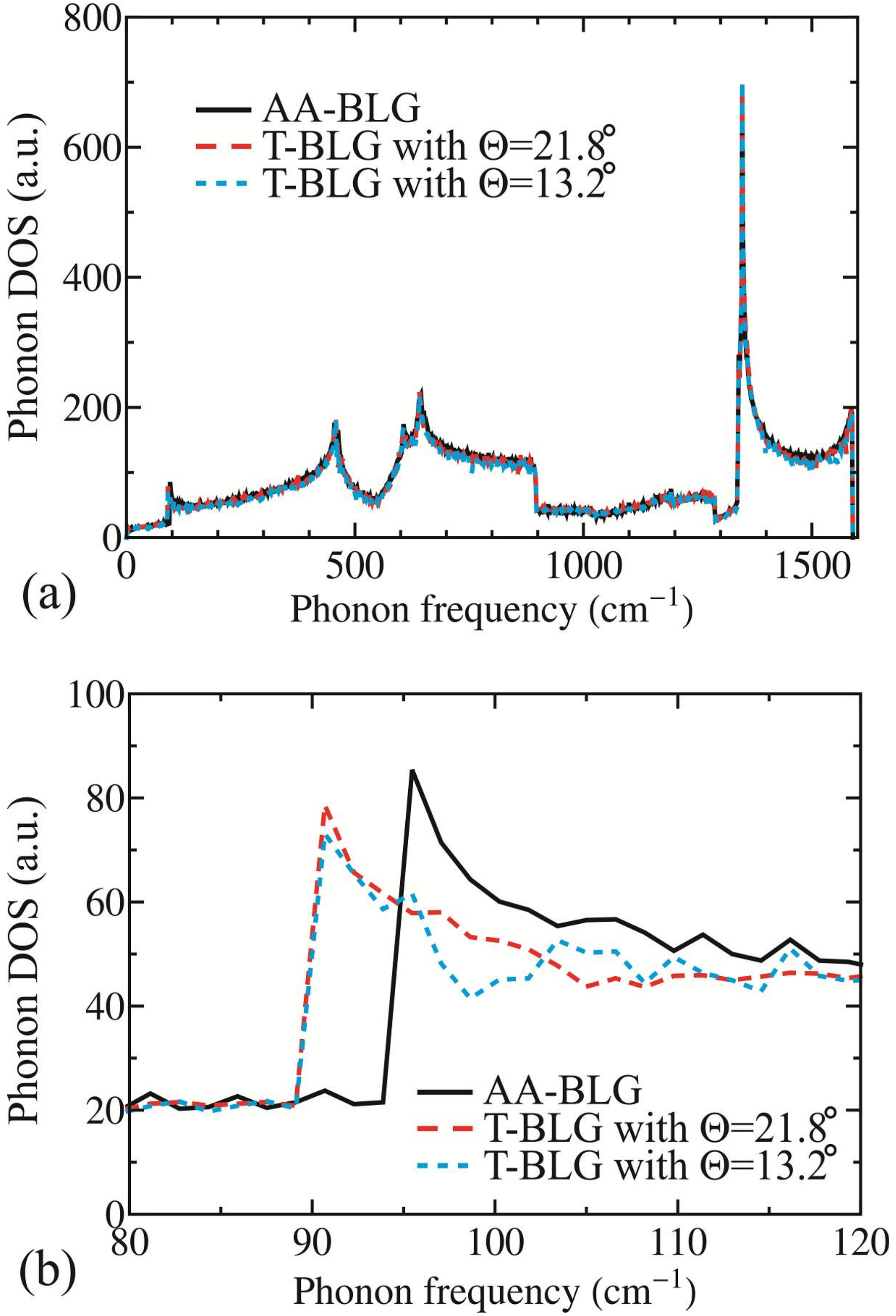

Figure 7

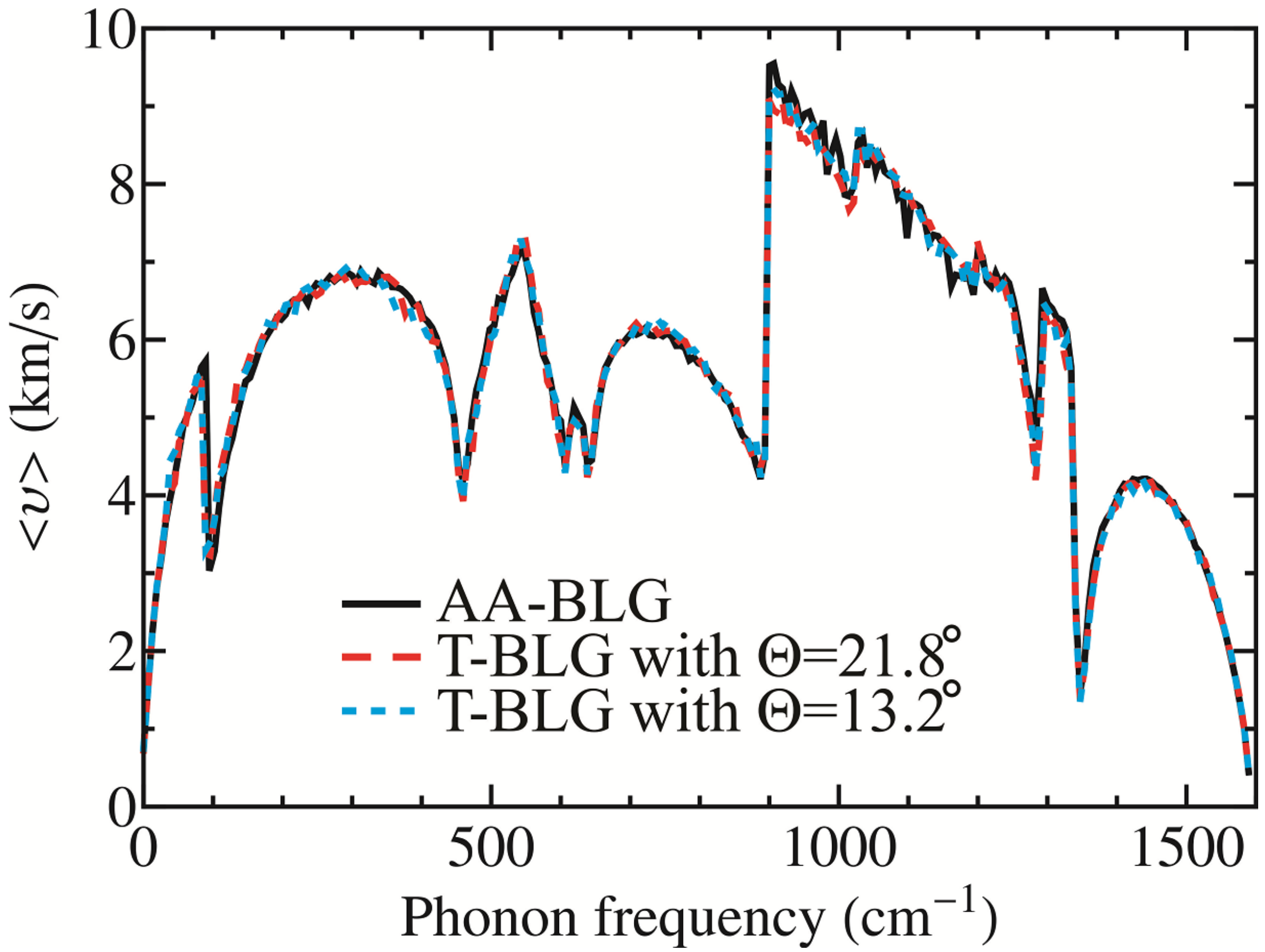

Figure 8